\documentclass[a4paper,11pt]{article}
\usepackage{pos}

\usepackage{dsfont}
\usepackage[english]{babel}
\usepackage[utf8]{inputenc}
\usepackage[T1]{fontenc}
\usepackage{tikz}

\title{Super-Higher-Form Symmetries}

\author[a,b]{Pietro Antonio Grassi}
\author*[c]{Silvia Penati}

\affiliation[a]{DiSIT, Universit\`a del Piemonte Orientale,\\ viale T. Michel, 11, 15121 Alessandria, Italy}

\affiliation[b]{INFN, Sezione di Torino, \\ via P. Giuria, 1, 10125, Torino, Italy}

\affiliation[c]{Dipartimento di Fisica, Universit\`a degli Studi di Milano–Bicocca and INFN, Sezione di Milano–Bicocca,\\
Piazza della Scienza 3, 20126 Milano, Italy}

\emailAdd{pietro.grassi@uniupo.it}
\emailAdd{silvia.penati@unimib.it}

\abstract{
We review the construction of higher-form symmetries for supersymmetric theories using a supergeometry framework. This reveals an enlarged set of topological conserved supercurrents, including supersymmetric Chern–Weil symmetries and new geometric Chern–Weil symmetries built from invariant supermanifold forms. In N=1 super-Maxwell theory in three dimensions, we construct the corresponding operators and charged defects, with charges determined by a super-linking number between their supporting hypersurfaces. 
At the end, we provide as an original unpubblished constribution some hints on how to construct super-symTFT for Chern-Weil and geometric Chern-Weil symmetries directly from supergravity.}

\FullConference{Proceedings of the Corfu Summer Institute 2025 "School and Workshops on Elementary Particle Physics and Gravity" and
Cost Action CaLISTA General Meeting 2025, ``Cartan, Generalised and Noncommutative Geometries, Lie Theory and Integrable Systems Meet Vision and Physical Models'', Corfu Summer Institute 2025}

\begin{document}
\maketitle

\section{Introduction}

Global symmetries play a crucial role in quantum field theory (QFT). They give rise to Ward identities
that constrain scattering amplitudes, allow to organize observables in group
representations, and their ’t Hooft anomalies constrain the RG flows and the strong coupling regime.

Higher-form symmetries have recently emerged as a new paradigm for investigating QFTs, also away from the perturbative regime. In fact, together with ordinary global symmetries and their anomalies, they open a new window on the infrared phases of a theory. They also provide a powerful tool for studying non-lagrangian theories and may play an important role in holography. Originally introduced by Kalb and Ramond in the string theory context \cite{Kalb:1974yc}, they have been largely studied in lattice field theory \cite{Wegner:1971app,
Savit:1977fw} and QFT \cite{Teitelboim:1985ya,Teitelboim:1985yc,Nambu:1975ba,Orland:1981ku}
in the eighties. More recently, they have been resumed by
Gaiotto, Kapustin, Seiberg, Willett \cite{Gaiotto:2014kfa}, with the goal of 
developing a
new framework to think about symmetries in QFT, where topological and
non-topological defects play an instrumental role.

In the modern language, a global $p$-form symmetry corresponds to the existence of a topological $p$-extended operator, which in turn is given by a conserved $(p+1)$-form current integrated on a codimension-$(p+1)$ smooth hypersurface. For $p=0$ we recover the usual Noether formulation of symmetries as being associated to conserved charges. Objects charged under a $p$-form symmetry are (non)topological $p$-extended objects, like Wilson loops for 1-form symmetries, Wilson surfaces for 2-form symmetries and Wilson-like hypersurfaces for $p>2$. Therefore,  the study of higher-form symmetries is strictly connected with the study of defect QFTs.

Higher-form symmetries exhibit several properties, beyond the obvious generalization of properties of ordinary symmetries. In fact, while they can have anomalies, they can be spontaneously broken, the non-anomalous ones can be gauged, as a distinguishing property they can be invertible or non-invertible \cite{Frohlich:2004ef} (see for instance \cite{SchaferNameki:2024ICTP,Shao:2023gho} for a review on recent developments). Moreover, for $p>0$ they are necessarily abelian symmetries, being the corresponding charges localized on non-space-filling hypersurfaces. 

Most of the results obtained so far restrict to bosonic higher-form symmetries, that is symmetries generated by bosonic  conserved currents. Only a few attempts have been made \cite{Chang:2022hud,Wang:2023iqt, Ambrosino:2024FermionicGeneralized,Bhardwaj:2024ydc,Huang:2024ror,Wen:2024udn} to generalize this construction to symmetries generated by spinorial currents. 
The most prominent example of fermionic symmetry is supersymmetry. It is then compelling trying to generalize the construction of higher-form symmetries to supersymmetric theories. We expect that combining higher-form symmetries with supersymmetry leads
to a new unexplored net of higher-form conservation laws. One then needs to
study their physical meaning and construct operators charged under these
new generators.

This program has been initiated in \cite{Grassi:2025tfp}, where supersymmetric theories have been naturally formulated in supermanifolds. This allows to exploit the language of supergeometry to treat bosonic and fermionic currents on an equal footing. Further progress has been recently obtained thanks to the construction of the associated Supersymmetric Symmetry TFT (SuSymTFT) \cite{Ambrosino:2026hhv}, which generalizes the original SymTFT construction to a supersymmetric enviromnent.

This contribution is primarily based on the content of \cite{Grassi:2025tfp}, with some input from \cite{Cremonini:2020mrk,Cremonini:2020zyn}. 
In the first section we briefly review the construction of ordinary higher-form symmetries, focusing on continuous, global, invertible symmetries. We will use the $n$-dimensional Maxwell theory as a probe. In section \ref{sect:susy} we present our construction of super-higher-form symmetries, still focusing on super-Maxwell theory in various dimensions and with different amount of supersymmetry. In section \ref{gCW} a brand new set of symmetries will be introduced, which we dub geometric-Chern-Weil symmetries. Finally, we devote section \ref{sect:general} to the collection of a few unpublished results concerning the supergravity derivation of the Symmetry Topological Field Theory (SymTFT)\footnote{See for instance \cite{Schafer-Nameki:2023jdn} for an introduction to the symTFT.} for supersymmetric theories. We conclude in section \ref{conclusions} with a short collection of possible follows-up. 

For all technical details and an exhaustive list of references we refer to \cite{Grassi:2025tfp}.

\section{Higher-Form Global Symmetries}

We begin by reviewing the construction of higher-form global symmetry generators for a $n$-dimensional field theory defined on a smooth manifold ${\cal M}^{(n)}$.

In geometric language, the theory exhibits a  $p$-form symmetry if it possesses a $(p+1)$-form current $J^{(p+1)}$ which on-shell  satisfies the conservation equation $d \star J^{(p+1)} \equiv d J^{(n-p-1)} = 0$. The corresponding generator is given by 
\begin{equation}
\label{eq:gen}
Q(\Sigma_{n-p-1})  = \int_{\Sigma_{n-p-1}} J^{(n-p-1)} = \int_{\mathcal M^{(n)}} J^{(n-p-1)} \wedge \mathbb{Y}_{\Sigma_{n-p-1}}^{(p+1)} \, ,
\end{equation}
where $\Sigma_{n-p-1}$ is a $(n-p-1)$-dimensional hypersurface and $\mathbb{Y}_{\Sigma_{n-p-1}}^{(p+1)}$ the corresponding Poincar\'e dual (alternatively dubbed Picture Changing Operator (PCO)), that is the delta-function operator that localizes the integral on $\Sigma_{n-p-1}$. The $Q$ operator is manifestly topological, thanks to the closure of the current and the fact that a change in the $\Sigma$-cycle corresponds to a shift of the PCO by a d-exact term.

\begin{figure}[ht]
    \centering \includegraphics[width=.3\textwidth]{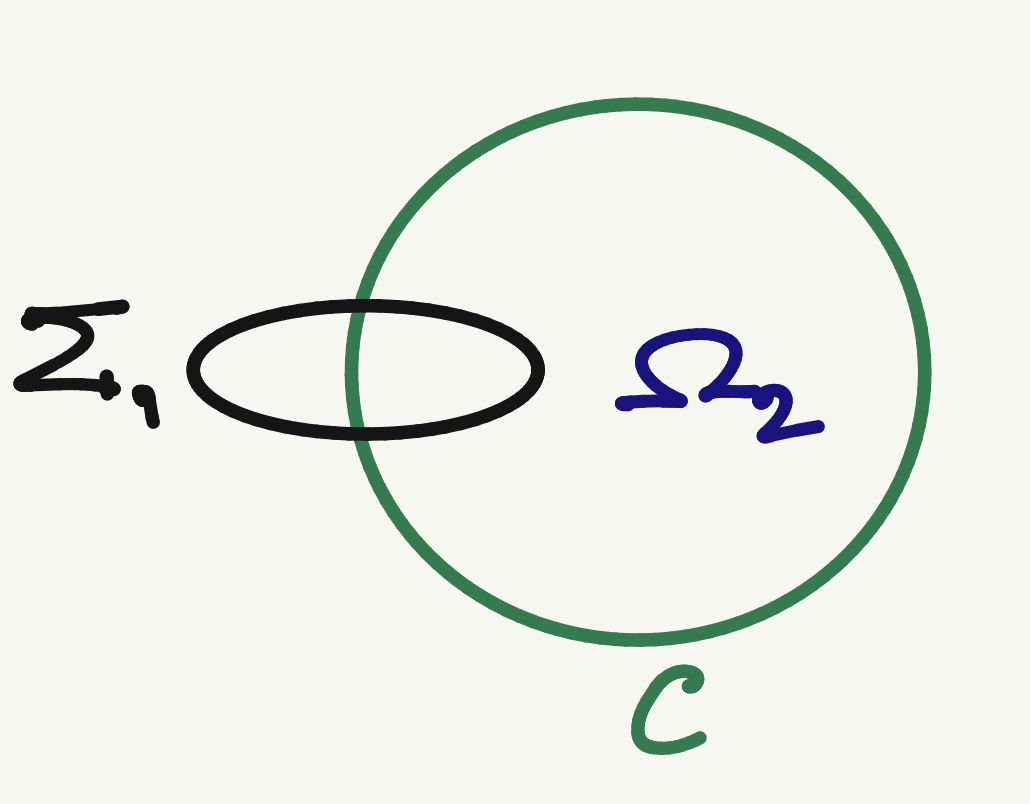}
    \caption{Two linked cycles in three dimensions.}
    \label{figure1}
\end{figure}

Operators charged under this symmetry are $p$-dimensional defects of the form
\begin{equation}
\label{eq:Wq}
    W(\Sigma_p) = e^{iq \Gamma} \, , \qquad \Gamma = \int_{\Sigma_p} A^{(p)} = \int_{{\mathcal M}^{(n)}} A^{(p)} \wedge \mathbb{Y}_{\Sigma_p}^{(n-p)}\, , 
\end{equation}
for a given $p$-form connection $A^{(p)}$, supported on a compact oriented $p$-dimensional submanifold $\Sigma_p$. In canonical formalism, the action of the symmetry generator on $W$ is given by 
\begin{equation}
\label{eq:QonW}
    e^{i\alpha Q(\Sigma_{n-p-1}) } \, W(\Sigma_p) \, e^{-i\alpha Q(\Sigma_{n-p-1}) }  = e^{i\alpha \, q \, {\rm Link}(\Sigma_{n-p-1}, \Sigma_p)} \, W(\Sigma_p)\, ,  
\end{equation}
where 
\begin{equation}
\label{eq:link}
{\rm Link}(\Sigma_{n-p-1}, \Sigma_p) \equiv 
\int_{{\mathcal M}^{(n)}}   \mathbb{Y}_{\Sigma_{n-p-1}}^{(p+1)}
\wedge \Theta_{\Omega_{p+1}}^{(n-p-1)} = (-1)^{p+1} \int_{{\mathcal M}^{(n)}}  \Theta_{\Omega_{n-p}}^{(p)} 
\wedge \mathbb{Y}_{\Sigma_p}^{(n-p)}    
\end{equation}
is the linking number between the two cycles \cite{botttu}, 
with $\partial \Omega_{p+1} = \Sigma_p$ and  $\partial \Omega_{n-p} = \Sigma_{n-p-1}$. Link is non-vanishing only if the dimensions of two hypersurfaces sum up to $(n-1)$ and the supports of the two integrated Poincar\'e duals in \eqref{eq:link}  have a non-trivial intersection, equivalently if the two surfaces $\Sigma_{n-p-1}$ and $\Sigma_p$ are disjoint. For the $n=3$ case, a simple representation of linked cycles is given in figure \ref{figure1}. 

As an illustrative example we consider Maxwell theory in $n$-dimensions. It is described by a 1-form potential $A$ whose dynamics is encoded into the action 
\begin{equation}
\label{eq:maction}
    S  = - \frac12 \int_{\mathcal M^{(n)}} F \wedge \star F \, , \qquad {\rm with} \quad F = dA \, .
\end{equation}
The equation of motion $d \star F=0$ and the Bianchi identity $dF=0$ provide two conservation laws. The  corresponding conserved charges 
\begin{equation}
\label{eq:Q}
    Q_e (\Sigma_{n-2})=  \int_{\mathcal M^{(n)}} \star F \wedge \mathbb{Y}_{\Sigma_{n-2}}^{(2)} \, , \qquad  \quad 
    Q_m (\Sigma_2) = \int_{\mathcal M^{(n)}} F \wedge \mathbb{Y}_{\Sigma_2}^{(n-2)} 
\end{equation}
are the well known generators of the 1-form electric and $(n-3)$-form magnetic $U(1)$ symmetries, respectively.

For manifolds with dimension $n > 2k$, further higher-form conserved currents can be constructed by taking wedge products of $F$,
\begin{equation}
J^{(2k)} = \frac{1}{k!} \, \underbrace{F \wedge F \wedge \dots F}_k \, .
\end{equation}
These are named Chern-Weil currents and have been discussed in \cite{Heidenreich:2020pkc, Brauner:2020rtz, Nakajima:2022feg}. The corresponding charges 
\begin{eqnarray}
    \label{CWB}
    Q^{(k)} (\Sigma_{2k})=\int_{\mathcal M} J^{(2k)} \wedge \mathbb{Y}^{(n-2k)}_{\Sigma_{2k}} 
\end{eqnarray}
generate $(n-2k-1)$-form U(1) global symmetries. 

Going back to the electric generator $Q_e$, the corresponding charged observables are 1-dimensional operators, that is ordinary Wilson loops of the form
\begin{equation}
    \label{GcwE}   
    W_q(\mathcal{C}) = \exp{\left( i q \int A \wedge \mathbb{Y}^{(n-1)}_{\mathcal{C}}\right) } = \exp{\left( i q \int F \wedge \Theta_{\Omega_2}^{(n-2)}\right) } \, ,
\end{equation}
where $\Omega_2$ is a two-dimensional surface whose boundary is the ${\cal C}$ cycle and $q$ is a real charge.

It is instructive to infer the corresponding charge under the action of the electric generator $U_e(\alpha, \Sigma_{n-2}) = e^{i \alpha Q_e(\Sigma_{n-2})}$. We will use a functional approach. To this end, we minimally couple the electric and magnetic currents to external background fields and consider the total action (up to a pure background term which does not have any effect)
\begin{equation}
\label{GcwB0}
    S + S_{min} =  \int_{{\mathcal M^{(n)}}} \left( - \frac12 (F - B^{(2)}) \wedge \star (F - B^{(2)}) +  F \wedge B^{(n-2)}  \right) \, .
\end{equation}
This action is invariant under the following gauge transformations
\begin{equation}
\label{eq:transfs1}
    F \rightarrow F + d\lambda^{(1)} 
    \, , \qquad 
   B^{(2)} \rightarrow B^{(2)} + d \lambda^{(1)} \, , \qquad
    B^{(n-2)} \rightarrow B^{(n-2)} + d\lambda^{(n-3)} \, ,
\end{equation}
up to a 't Hooft anomaly term, precisely
\begin{equation}
\label{eq:anomaly0}
S' + S_{min}' = S + S_{min} - \int_{\mathcal M^{(n)}}   \lambda^{(1)} \wedge d B^{(n-2)} \, .
\end{equation} 
Note that the gauge transformation of $B^{(n-2)}$ 
has gauge-for-gauge symmetries such as $\lambda^{(n-3)} \rightarrow \lambda^{(n-3)} + d \lambda^{(n-4)}$, and similarly for $\lambda^{(n-4)}$ if 
$n> 3$. This is crucial for implementing the topological symmetries of the theory. 

Upon defining the functional integral 
\begin{equation}
     \label{GcwC0}
      \mathbb{Z}[B^{(2)}, B^{(n-2)}] =\int [\mathcal{D} A] \, e^{ i \, ( S +S_{min} + S_{gf})} \, ,
 \end{equation}
with a suitable gauge fixing term $S_{gf}$,
the electric charge of the Wilson loop can be computed from the two-point function 
$\Big\langle U_e(\alpha, \Sigma_{n-2})\, W_q(\mathcal{C}) \Big\rangle$. On the other hand, it is easy to realize that the insertion of $U_e$ in the functional integral corresponds to turning on a background gauge field $B^{(2)} = \alpha \mathbb{Y}^{(2)}_{\Sigma_{n-2}}$. Analogously, the insertion of $W_q(\mathcal{C})$ corresponds to turning on $B^{(n-2)} = q \Theta^{(n-2)}_{\Omega_2}$. Therefore, we have
$\Big\langle U_e(\alpha, \Sigma_{n-2})\, W_q(\mathcal{C}) \Big\rangle = \mathbb{Z}[\alpha \mathbb{Y}^{(2)}_{\Sigma_{n-2}}, q \Theta^{(n-2)}_{\Omega_2}]$, with $\mathbb{Z}$ defined in \eqref{GcwC0}.  We can now gauge away $B^{(2)}$ by performing a gauge  transformation \eqref{eq:transfs1} with $d\lambda^{(1)} = - \alpha \mathbb{Y}^{(2)}_{\Sigma_{n-2}}$, thus obtaining $Z[0, q \Theta^{(n-2)}_{\Omega_2}] = \langle W_q(\mathcal{C}) \rangle$. However, as mentioned above, this gauge transformation is anomalous. Therefore, taking into account the explicit expression \eqref{eq:anomaly0} for the anomaly and inserting there the specific choices of $\lambda^{(1)}$ and $B^{(n-2)}$, we eventually find 
\begin{equation}
    \label{GcwF2}
    \Big\langle U_e(\alpha, \Sigma_{n-2}) \, W(\mathcal{C}) \Big\rangle  = e^{i \alpha q \int  \mathbb{Y}^{(2)}_{\Sigma_{n-2}} \wedge \Theta^{(n-2)}_{\Omega_2}} \, \Big\langle W(\mathcal{C})\Big\rangle \, .
\end{equation} 
The Wilson loop is then charged under the electric $U(1)$ Noether-like 1-form symmetry, with charge proportional to the linking number ${\rm Link}({\mathcal C}, \Sigma_{n-2}) = \int  \mathbb{Y}^{(2)}_{\Sigma_{n-2}}\wedge \Theta^{(n-2)}_{\Omega_2}$. It is important to stress that the charge is entirely due to the 't Hooft anomaly \eqref{eq:anomaly0}. 

In a similar manner it is possible to prove that the observable charged under the action of the magnetic generator $U_m(\alpha_m, \Sigma_2) = e^{ia_m Q_m(\Sigma_2)}$ is an extended monopole supported on a 't Hooft surface $\Sigma_{n-3} \subset {\mathcal M}^{(n)}$, which sources a singular field strength $ dF = q_m \mathbb{Y}^{(3)}_{\Sigma_{n-3}}$. The corresponding charge is $\alpha_m q_m {\rm Link}(\Sigma_2, \Omega_{n-2})$ where $\Sigma_2$ supports the $Q_m$ generator and $\partial \Omega_{n-2} = \Sigma_{n-3}$. Still, the charge is entirely due to a 't Hooft anomaly.

A similar analysis can be applied to the Chern-Weil charges defined in \eqref{CWB}. The corresponding charged operators can be constructed from non-trivial intersections of extended monopoles. This has been discussed in \cite{Nakajima:2022feg} and we refer to this paper for more details.

\section{Adding Supersymmetry}
\label{sect:susy}

In this section we generalize the previous construction to theories where the spectrum of symmetries includes supersymmetry. 

Supersymmetric theories can be conveniently formulated in supermanifolds (see for instance \cite{Dewitt,Rogers}), where supersymmetry invariance is made manifest. The natural objects that live in a supermanifold are pseudoforms, superforms and integral forms. Here we briefly review their definition, together with the differential and  integral calculus in supermanifolds, and then exploit this framework to construct super-higher-form symmetries for supersymmetric theories.  

We consider a field theory defined in the supermanifold $\mathcal{SM}^{(n|m)}$,
where $n$ is the bosonic dimension and $m$ is the fermionic one ($m$ equals the number of supercharges). Locally, 
it is isomorphic to a superspace $\mathbb{R}^{(n|m)}$ parametrized by coordinates $z^A = (x^a, \theta^\alpha)$, where $a=0,\dots,n-1, \alpha=1,\dots,m$. In the table below we report how vielbeins, differential and forms are generalized in the presence of odd coordinates,  

\begin{center}
\begin{tabular}{c|c|c}
~~  & ${\cal M}^{(n)}$ & ${\cal SM}^{(n|m)}$ \\
\hline
~ & ~~ & ~ \\
\hspace{-0.5cm} {Coordinates} & $x^a$ & $(x^a, \theta^\alpha)$ \\
~ & ~~ & ~ \\
\hspace{-0.5cm}(Super)vielbeins & $V^a = dx^a$  & \hspace{-0.2cm} $V^a = dx^a + \theta^\alpha \gamma^a_{\alpha\beta} d\theta^\beta , \; \; \psi^\alpha = d\theta^\alpha  $ \\
~ & ~~ & ~ \\
\hspace{-0.5cm} Differential & $d = V^a \partial_a $ & $d = V^a \partial_a + \psi^\alpha D_\alpha$ \\
~ & ~~ & ~ \\
\hspace{-0.5cm}Forms & $J^{(p)} = J_{[a_1 ... a_p]}(x)  V^{a_1} ... V^{a_p}$ & $J^{(p|q)} \quad 0 \leq q \leq m$ \\
~ & ~~ & ~ \\
 ~ & ~~ & ~ \\
\end{tabular} 
\end{center}  

\noindent
Here, $D_\alpha$ are the supersymmetry covariant derivatives in superspace. 
Moreover, $J^{(p|q)}$ indicates a generic pseudoform with form degree $p$ and picture $q$.\footnote{We use notations and conventions of \cite{Cremonini:2020mrk,Cremonini:2020zyn}, where more details on the construction of pseudoforms in supermanifolds can be found.}  
In this presentation  we will be primarily interested in the two extreme cases, $q=0$ (superforms) and $q=m$ (integral forms). They have the following general expansion in powers of the supervielbeins 
\begin{eqnarray}
J^{(p|0)} &=& \sum_{k=0}^p J_{[a_1 \dots a_k](\alpha_{k+1} \dots \alpha_{p})}(x,\theta) \; V^{a_1} \dots V^{a_k}  
    \psi^{\alpha_{k+1}} \dots \psi^{\alpha_p}  \, ,\nonumber \\
J^{(p|m)} &=& 
 \sum_{k=p}^n J_{[a_1 \dots a_k]}^{(\alpha_{1} \dots \alpha_{k-p})}(x,\theta) \;  V^{a_1} \dots V^{a_k} 
    \iota_{\alpha_{1}} \dots \iota_{\alpha_{k-p}}  \delta^{(m)}(\psi)  \, , 
\end{eqnarray}
where the coefficients are ordinary superfields with non-trivial tensiorial and/or spinorial nature. 

It is important to stress that, while the form degree is kept fixed as in the ordinary manifolds, the picture can be modified by acting on the pseudoforms with suitable rising or lowering super-PCO (see for example appendix A of \cite{Ambrosino:2026hhv} for a short review). 

Integration theory on supermanifolds have been developed in \cite{Witten:2012bg}. This requires introducing an invariant super-measure $[dx d(dx) d\theta d(d\theta)]$ subject to the following defining conditions
\begin{equation}
    \int [dx] dx \equiv \int [dx]  \delta(dx) = 1, \qquad \int d[d\theta] \delta (d \theta) = 1 \, ,
\end{equation}
such that the integral of a top (integral) form is given by
\begin{eqnarray}
    \int \! J^{(n|m)} &\equiv& \int_{{\cal SM}^{(n|m)}}  [dx d(dx)  d\theta  d(d\theta)] \, J_{[a_1 ... a_n]}(x, \theta)  \, V^{a_1} ... V^{a_n} \delta^{(m)}(d\theta) \nonumber \\
&=& \int_{{\cal SM}^{(n|m)}}  [dx d\theta] \, J_{[a_1 ... a_n]}(x,\theta) \, .
\end{eqnarray}
The expression in the second line is nothing but the usual Berezin integral in superspace. 

Integration can be generalized to lower-degree pseudoforms integrated on submanifolds, as 
\begin{equation}
  \int_{\Sigma_{(p|0)}} J^{(p|0)} = \int_{{\cal SM}^{(n|m)}} J^{(p|0)}  \wedge {\mathbb Y}_{\Sigma_{(p|0)}}^{(n-p|m)} \qquad \quad \int_{\Sigma_{(p|m)}} J^{(p|m)} = \int_{{\cal SM}^{(n|m)}} J^{(p|m)}  \wedge {\mathbb Y}_{\Sigma_{(p|m)}}^{(n-p|0)} \, ,
\end{equation}
where ${\mathbb Y}_{\Sigma_{(p|0)}}^{(n-p|m)}$ and ${\mathbb Y}_{\Sigma_{(p|m)}}^{(n-p|0)}$ are the PCOs that project the integrals on the corresponding submanifolds. 

Armed with this preliminary set of information, we are now ready to construct $(p|0)$-superform and $(p|m)$-integral form symmetries by generalizing definition \eqref{eq:gen} for ordinary manifolds. 

\begin{itemize}
\item A $(p|0)$-superform symmetry is generated by a closed $(n-p-1|m)$  current, or equivalently, a conserved $(p+1|0)$-superform current, that is $ d J^{(n-p-1|m)} \equiv d \star J^{(p+1|0)} =  0$. The corresponding conserved charge
\begin{equation}
    Q(\Sigma_{(n-p-1|m)}) =\int_{{\cal SM}^{(n|m)}} J^{(n-p-1|m)}\wedge \mathbb{Y}^{(p+1|0)}_{\Sigma_{(n-p-1|m)}}
\end{equation}
is a topological operator. In fact, a smooth change of the $\Sigma_{(n-p-1|m)}$ cycle corresponds to a shift of the PCO by a $d$-exact term, which in the integral gives a vanishing contribution due to the closure of the current. This type of currents are the supersymmetric generalization of Noether currents. 

Objects charged under these symmetries are generalized Wilson-like operators in supermanifold \cite{Cremonini:2020mrk,Cremonini:2020zyn}. Precisely, these are operators given by the holonomy of a $(p|0)$-superconnection, localized on a $(p|0)$-cycle 
\begin{equation}
    W_q(\Sigma_{(p|0)}) = \exp \left( i q \int_{\Sigma_{(p|0)}} A^{(p|0)} \right) = \exp \left( i q \int_{{\cal SM}^{(n|m)}} A^{(p|0)} \wedge \mathbb{Y}^{(n-p|m)}_{\Sigma_{(p|0)}} \right) \, .
\end{equation}

The action of the (super)symmetry generator $U_\alpha(\Sigma_{(n-p-1|m)}) = e^{i\alpha \, Q(\Sigma_{(n-p-1|m)})}$ on $W$ can be easily figured out by generalizing the construction of section 3 to the supermanifold. Formally, the procedure is very similar and one eventually obtains (below, we report details for ${\cal N}=1$ super-Maxwell theory in three dimensions)
\begin{equation}
    \langle e^{i\alpha \, Q(\Sigma_{(n-p-1|m)})} \, W_q (\Sigma_{(p|0)}) \rangle  = e^{i\alpha q\, {\rm SLink}\left(\Sigma_{(p|0)},   \Sigma_{(n-p-1|m)}\right) } \langle W_q (\Sigma_{(p|0)})\rangle \, ,
\end{equation}
where 
\begin{equation}
    {\rm SLink}\left(\Sigma_{(p|0)},  \Sigma_{(n-p-1|m)}\right)  
 =  \int_{{\cal SM}^{(n|m)}}  \! \Theta^{(n-p-1|m)}_{\Omega_{(p+1|0)}} \wedge  \mathbb{Y}^{(p+1|0)}_{\Sigma_{(n-p-1|m)}} \, , \qquad \partial \Omega_{(p+1|0)} = \Sigma_{(p|0)}
\end{equation}
is the superlinking number which measures the intertwining of the two supercycles in supermanifold \cite{Cremonini:2020zyn}. It is interesting to note that it is non-vanishing also when the two cycles wrap only in the odd directions. 

\vskip 5pt
\item 
A $(p|m)$-integral form symmetry is generated by a closed $(n-p-1|0)$  current, or equivalently, a conserved $(p+1|m)$-superform current, that is $ d J^{(n-p-1|0)} \equiv d \star J^{(p+1|m)} =  0$. Similarly to the previous case, the corresponding conserved charge
\begin{equation}
    Q(\Sigma_{(n-p-1|0)}) =\int_{{\cal SM}^{(n|m)}}  J^{(n-p-1|0)}\wedge \mathbb{Y}^{(p+1|m)}_{\Sigma_{(n-p-1|0)}}
\end{equation}
is a topological operator, thanks to the closure of the current. This type of currents are the supersymmetric generalization of topological currents. 

Therefore, the corresponding charged objects are higher-dimensional ’t Hooft-like {\em super-monopoles} supported on a singular hypersurface $\Sigma_{(p|m)}$, which  affects the closure of $J^{(n-p-1|0)}$ by a localized term, $d J^{(n-p-1|0)} = 
\tilde{q} {\mathbb Y}^{(n-p|0)}_{\Sigma_{(p|m)}}$. The corresponding charge is still given by the superlinking number 
\begin{equation}
    {\rm SLink}\left(\Sigma_{(p|m)},  \Sigma_{(n-p-1|0)}\right)  
=  \int_{{\cal SM}^{(n|m)}} \Theta^{(n-p-1|0)}_{\Omega_{(p+1|m)}} \wedge  \mathbb{Y}^{(p+1|m)}_{\Sigma_{(n-p-1|0)}} \, , \qquad \partial \Omega_{(p+1|m)} = \Sigma_{(p|m)} \, .
\end{equation}
\end{itemize}

\noindent
More generally, one could consider $(p|q)$-form symmetries with $q \neq 0,m$. These are generated by closed $(n-p-1|m-q)$-form currents and represent a very rich sector of new symmetries. Nonetheless, defining these pseudo-form currents 
requires a different geometric framework. We plan to construct them in the future.

Independently of their explicit construction, a natural question that can be addressed is whether the topological conserved charges generating super-higher-form symmetries,
\begin{equation}
    Q(\Sigma_{(n-p-1|m-q)}) = \int_{\Sigma_{(n-p-1|m-q)}} J^{(n-p-1|m-q)} \, , \qquad 0 \leq q \leq m
\end{equation}
are abelian or non-abelian. Applying a reasoning similar to the one for bosonic higher-form symmetry charges, we see that if $p=q=0$, the integral is on a superspace filling surface $\Sigma_{(n-1|m)}$ (only the time direction is left out), so they can be non-commuting, since there is no extra even or odd  dimension
along which we can exchange $Q(t - \epsilon)$ with $Q(t + \epsilon)$. Instead, if $p$ and/or $q$ are different from zero the surface is no longer superspace filling, thus there are even and/or odd directions along which we can exchange the two charges. We note that this is true also for surfaces that fill all the spatial bosonic dimensions but do not fill all the fermionic ones. Therefore, $(0|q)$-form symmetries with $q \neq 0$ are all abelian, whereas only $(0|0)$-form symmetries can be non-abelian.

\vskip 5pt

In order to make this  construction more concrete, we specialize it to the super-Maxwell theory in $n$ dimensions with $m$ supercharges. 
In ${\cal SM}^{(n|m)}$ supermanifold, it is described by a $(1|0)$-form superpotential $A = A_a dx^a + A_\alpha \psi^\alpha$. Its field strength $F= dA$ is the $(2|0)$-superform 
\begin{equation}
\label{Fexpansion}
    F = F_{ab} V^a \wedge V^b + F_{a\alpha} V^a \wedge \psi^\alpha + F_{\alpha\beta} \, \psi^\alpha \wedge \psi^\beta \, ,
\end{equation}
where $F_{ab}, F_{\alpha\beta}$ and $W^\beta$ are the field strength and the gluino superfields, respectively. It is invariant under supergauge transformations, 
$\delta A = d \Lambda$, and satisfies the Bianchi identity $d F = 0$.  Redundant degrees of freedom in $F$ are removed by imposing suitable gauge invariant constraints (for example, the conventional constraint $F_{\alpha\beta} = 0$). 

The super-Maxwell action reads 
\begin{equation}
\label{eq:superaction}
    S = -\frac12\int_{{\cal SM}^{(n|m)}}F \wedge \star F \, , 
\end{equation}
and provides the equation of motion $d \star F = 0$. The equation of motion together with the Bianchi identity easily allow to construct closed supercurrents:

\noindent 
\begin{itemize}
    \item The Noether-like current $J^{(n-2|m)} = \star F$
    \item The topological currents $ J^{(2k|0)} = \frac{1}{k!} \, \underbrace{F \wedge F \wedge \dots F}_k\, , n >2k$. These are dubbed super-Chern-Weil currents.
\end{itemize}

\noindent
The corresponding conserved charges can be obtained by integrating these currents on suitable super-cycles, precisely 
\begin{equation}
    Q(\Sigma_{(n-2,m)}) = \int_{{\cal SM}^{(n|m)}} \! \! J^{(n-2|m)} \wedge \mathbb{Y}^{(2|0)}_{\Sigma_{(n-2,m)}} \, , \qquad Q_{\rm CW} (\Sigma_{(2k,0)}) = \int_{{\cal SM}^{(n|m)}} \! \! J^{(2k|0)} \wedge \mathbb{Y}^{(n-2k|m)}_{\Sigma_{(2k,0)}} \, .
\end{equation}
They generate a $(1|0)$-superform symmetry and $(n-2k-1,m)$-integral form symmetries, respectively. They are topological operators, thanks to the closure of the currents and the behavior of the PCO under a smooth deformation of the corresponding super-cycle. 

In \cite{Grassi:2025tfp} we have explicitly studied super-Maxwell theory for $(n|m) = (3|2), (4|4), (6|8), (10|16)$. As an illustrative example, here we report the $(3|2)$ case that corresponds to ${\cal N}=1$ super-Maxwell theory in three dimensions. The $\mathcal{SM}^{(3|2)}$ supermanifold is parametrized by local coordinates $x^a, a = 0,1,2$ and $\theta^\alpha, \alpha =1,2$.

In three dimensions the only closed currents that can be constructed are the Noether-like electric current $J^{(1|2)} = \star F$, and the topological magnetic current $J^{(2|0)} = F$. 
As follows from \eqref{Fexpansion} and the definition of super-Hodge dual in ${\cal SM}^{(3|2)}$ \cite{Castellani:2015paa,
Castellani:2015ata}, their expansions read
\begin{eqnarray}
J^{(1|2)} &=& F^{ab} \epsilon_{abc} V^c \wedge \delta^{(2)}(\psi) + 
W^\alpha \sigma^a_{\alpha \beta}\epsilon_{abc} V^b \wedge V^c \wedge \iota^\beta \delta^{(2)}(\psi) \, , \nonumber \\
J^{(2|0)} &=& F_{ab} V^a \wedge V^b + F_{a\alpha} V^a \wedge \psi^\alpha \, ,
\end{eqnarray}
where conventional constraints have been implemented, $\iota_\beta \equiv \partial/\partial \psi^\beta$ is the contraction along the odd directions and $\{ \sigma^a \}$ are the Pauli matrices. The  components superfields satisfy the on-shell conditions 
\begin{eqnarray}
\label{S3E}
\partial_{(\alpha}^{~~\beta} f_{\beta|\gamma)} =0\,,~~~~ \partial^{\alpha\beta}W_\beta =0 \, , ~~~~
\partial^{\alpha\beta} f_{\alpha\beta} =0 \, , \qquad {\rm where } \quad f_{\alpha\beta} \equiv (\sigma^{ab})_{\alpha\beta} F_{ab} \, .
\end{eqnarray}
Round brackets among some equal-type-of indices denote symmetrization of those indices. 
The corresponding generators $U_e(\Sigma_{(1|2)}, \alpha_e) = e^{i \alpha_e Q_e(\Sigma_{(1|2)})}$ and $U_m(\Sigma_{(2|0)}, \alpha_m) = e^{i \alpha_m Q_m(\Sigma_{(2|0)})}$ are written in terms of the two topological operators 
\begin{equation}
\label{S3F}
Q_e(\Sigma_{(1|2)}) = \int_{{\cal SM}^{(3|2)}} \! \!  \star F
\wedge \mathbb{Y}^{(2|0)}_{\Sigma_{(1|2)}}\, , \qquad \qquad
Q_m(\Sigma_{(2|0)}) =  \int_{{\cal SM}^{(3|2)}} \! \!  F \wedge \mathbb{Y}^{(1|2)}_{\Sigma_{(2|0)}} \, . 
\end{equation}
Here $\Sigma_{(1|2)}$ is a superline and  $\Sigma_{(2|0)}$ an ordinary surface in ${\cal SM}^{(3|2)}$. The two PCOs localize the integrals on these two super-cycles. 

In order to study operators charged under these symmetries and evaluate their charge it is convenient to couple the two currents to external backgrounds, a $(2|0)$-superform $B_e$ and a $(1|2)$-integral form $B_m$, and consider the following functional integral
\begin{equation}
\label{functional}
    {\mathbb Z}[B_e, B_m] = \int [{\cal D}A] e^{i(S + S_{min} +S_{gf})} \, ,
\end{equation}
with (we neglect the gauge fixing term) 
\begin{equation}
    S+ S_{min} = \int_{{\cal SM}^{(3|2)}} \! \! \left( - \frac12 (F - B_e) \wedge \star (F - B_e) +  F \wedge B_m \right) \, .
\end{equation}
Under gauge transformations
\begin{eqnarray}
\label{eq:transfs1A}
    F \rightarrow F + d\Lambda^{(1|0)} 
    \, , \qquad  B_e \rightarrow B_e + d \Lambda^{(1|0)} \, , \qquad 
    B_m \rightarrow B_m + d \Lambda^{(0|2)} \, ,
\end{eqnarray}
the action is invariant, up to an anomaly term 
\begin{equation}
\label{anomaly}
    S' + S_{min}' = S + S_{min} 
    - \int_{{\cal SM}^{(3|2)}} \! \!  \Lambda^{(1|0)} \wedge dB_m \, .
\end{equation}
Note, as above, the gauge symmetry of $B_m$, which is an integral form has some gauge-for-gauge symmetries such as $ \Lambda^{(0|2)} \rightarrow \Lambda^{(0|2)} + d \Lambda^{(-1|2)}$ and similarly $\Lambda^{(p|2)} \to \Lambda^{(p|2)} + d\Lambda^{(p-1|2)}, p < 0$. This gauge-for-gauge symmetries provide an infinite set of gauge parameters neeeded in order to implement the correct topological symmetries. The mathematical structure of these symmetries will be discussed in a forthcoming paper. 

Looking for operators that are charged under these symmetries, it is immediate to realize that super-Wilson loops of the form \cite{Cremonini:2020mrk} 
\begin{equation}
W_{q_e}(\mathcal{C}) = e^{i q_e \Gamma } \,, \quad {\rm with} \quad
\Gamma = \int_{{\cal SM}^{(3|2)}} \! \!  A \wedge \mathbb{Y}^{(2|2)}_\mathcal{C} = \int_{{\cal SM}^{(3|2)}} \! \!  F \wedge \Theta^{(1|2)}_{\Omega_{(2|0)}} \, , \;\; \quad \partial \Omega_{(2|0)} = \mathcal{C}
\end{equation}
are charged under $U_e$. Proceeding as in the non-supersymmetric case, that is computing the two-point function $\langle U_e W_{q_e} \rangle$ as the functional integral with a specific choice of the backgrounds, and exploit the anomalous gauge invariance \eqref{eq:transfs1A} to remove $U_e$, we eventually find 
\begin{equation}
  \Big\langle U_e(\Sigma_{(1|2)}, \alpha_e) \, W_{q_e}(\mathcal{C}) \Big\rangle =
    \mathbb{Z}[\alpha_e \mathbb{Y}^{(2|0)}_{\Sigma_{(1|2)}}, \, q_e \Theta^{(1|2)}_{\Omega_{(2|0)}}] = 
    e^{i \alpha_e q_e \, {\rm SLink}(\Sigma_{(1|2)}, \, \mathcal{C}) } \Big\langle W_{q_e}({\mathcal C}) \Big\rangle \, , 
 \end{equation}
 where the charge is given by the super-linking number
 \begin{equation}
{\rm SLink}(\Sigma_{(1|2)}, \, \mathcal{C})  = \int_{{\cal SM}^{(3|2)}} \! \! \mathbb{Y}^{(2|0)}_{\Sigma_{(1|2)}} \wedge \Theta^{(1|2)}_{\Omega_{(2|0)}} = 
 \int_{{\cal SM}^{(3|2)}} \! \! \Theta^{(1|0)}_{\Omega_{(2|2)}}\wedge \mathbb{Y}^{(2|2)}_{\mathcal{C}} \, .
\end{equation}
It is important to stress that the charge originates entirely from the anomaly in \eqref{anomaly}.

Analogously, we can determine the operators charged under $U_m$. These are point-like monopoles $P_{(0|2)}$ that source a singular term in the current conservation law, $dF = q_m \mathbb{Y}^{(3|0)}_{P_{(0|2)}} $. Once again, computing correlation functions with the generating functional \eqref{functional}, we find 
\begin{equation}
  \Big\langle   U(\Sigma_{(2|0)}, \alpha_m) \, P_{(0|2)} \Big\rangle = 
  \mathbb{Z} [q_m\Theta^{(2|0)}_{\Sigma_{(1|2)}} , \, \alpha_m \mathbb{Y}^{(1|2)}_{\Sigma_{(2|0)}}] = e^{i \alpha_m q_m \, {\rm SLink}(P_{(0|2)}, \Sigma_{(2|0)})} 
  \Big\langle P_{(0|2)} \Big\rangle \, ,
\end{equation} 
with the super-linking number 
\begin{equation}
   {\rm SLink}(P^{(0|2)}, \Sigma_{(2|0)}) = \int_{{\cal SM}^{(3|2)}} \! \!  \Theta^{(2|0)}_{\Sigma_{(1|2)}} \wedge \mathbb{Y}^{(1|2)}_{\Sigma_{(2|0)}} 
\end{equation}
entirely determined by the anomaly in \eqref{anomaly}. 

We complete the spectrum of symmetries of ${\cal N}=1$ Maxwell theory in three dimensions by constructing the  supersymmetry currents. 
We recall that in supermanifolds supersymmetry is realized linearly as coordinates (super)translations. These are 0-form symmetries and are generated by the following $(2|2)$-currents \cite{Grassi:2025tfp}
\begin{equation}
J^{(2|2)}_\alpha = (V \wedge V)^{\beta\gamma} f_{\alpha\beta}W_\gamma \,\delta^{(2)}(\psi)     \, , \qquad 
J^{(2|2)}_{(\alpha\beta)} = (V \wedge V)^{\gamma\delta}\left(  f_{\alpha\gamma} f_{\beta\delta} 
+ W_\gamma \partial_{  \alpha \beta} W_\delta
\right) \delta^{(2)}(\psi)  \, .
\end{equation}
Their closure
is ensured by the Bianchi identities in components, $D_\alpha f_{\beta\gamma} = 
\partial_{\alpha (\beta} W_{\gamma)}$ and 
$f_{\alpha \beta} = D_\alpha W_\beta$, 
and the component equations of motion \eqref{S3E}. Moreover, $\partial^{\alpha\beta} J^{(2|2)}_{(\alpha\beta)} =0$ on-shell. 
The superfield components of $J^{(2|2)}_\alpha$ and $J^{(2|2)}_{(\alpha\beta)}$ are the superfield version of the supersymmetry and stress-tensor currents, respectively \cite{Grassi:2025tfp}.

\section{Geometric super-Chern-Weil currents}
\label{gCW}

The spectrum of super-Chern-Weil currents that we have reviewed for Maxwell theory do not exhaust the entire spectrum of possible currents of this type. In fact, as discussed in \cite{Grassi:2025tfp}, we can exploit geometric objects like supervielbeins to build up new closed forms $\omega$ that can give rise to new closed currents of the form $\omega \wedge J$, where $J$ is another closed current of the theory. These were named geometric-Chern-Weil (gCW) symmetries. 

As a first illustrative example, we consider ${\cal N}=1$ super-Maxwell theory in four dimensions. It lives in ${\mathcal{SM}}^{(4|4)}$, parametrized by local coordinates $(x^{\alpha \dot\alpha}, \theta^\alpha, \bar\theta^{\dot\alpha})$, $\alpha =1,2, \dot\alpha = \dot{1},\dot{2}$. Given the supervielbeins $(V^{\alpha \dot\alpha} = dx^{\alpha\dot\alpha} + \theta^\alpha \bar\psi^{\dot \alpha} + \psi^{\alpha}
\bar \theta^{\dot \alpha} \,, 
\psi^{\alpha} = d \theta^\alpha, \bar\psi^{\dot\alpha} = d \bar\theta^{\dot\alpha})$ we can construct the following closed forms 
\begin{equation}
    \label{cocA}
    \omega^{(3|0)} = V^{\alpha\dot\alpha} \wedge \psi_\alpha \wedge \bar \psi_{\dot\alpha}\,, ~~~~~
    \omega^{(4|0)} = (V\wedge V)^{\alpha\beta} \wedge \psi_\alpha \wedge \psi_{\beta}\,, ~~~~~
    \bar\omega^{(4|0)} = (V\wedge V)^{\dot\alpha\dot\beta} \wedge\bar\psi_{\dot\alpha} \wedge \bar\psi_{\dot\beta}\, , 
\end{equation}
and the corresponding Hodge duals 
\begin{eqnarray}
    \label{cocB}
    &&\omega^{(1|4)} \equiv \star  \omega^{(3|0)} =  (V\wedge V\wedge V)^{\alpha\dot\alpha} \wedge \iota_\alpha \bar \iota_{\dot\alpha}  \, \delta^{(4)}(\psi)\,, ~~~~~ \nonumber \\
    &&\omega^{(0|4)} \equiv \star  \omega^{(4|0)} =(V\wedge V)^{\alpha\beta} \wedge \iota_\alpha \iota_{\beta}  \, \delta^{(4)}(\psi)\,, ~~~~~ \nonumber \\
     &&\bar\omega^{(0|4)}  \equiv \star  \bar\omega^{(4|0)} =(V\wedge V)^{\dot\alpha\dot\beta} \wedge \bar\iota_{\dot\alpha} \bar \iota_{\dot \beta} \, \delta^{(4)}(\psi)\, . 
\end{eqnarray}
Using these objects we can then define three additional conserved integral-form currents
\begin{equation}
\label{eq:omegacurrents}
    J^{(3|4)} = \omega^{(1|4)} \wedge F \, , \qquad 
    J^{(2|4)} = \omega^{(0|4)} \wedge F \, , \qquad 
    \bar{J}^{(2|4)} = \bar\omega^{(0|4)} \wedge F \, ,
\end{equation}
which allow to define topological symmetry generators, once we integrate them on cycles with the correct dimensions. 

This construction can be easily generalize to $(n|m)$ super-Maxwell theories for suitable even and odd dimensions. Higher is the bosonic dimension, larger is the spectrum of gCW currents that one can construct. They generate a new set of higher-form symmetries that would definitively deserve a deeper investigation.

As a second example, we discuss the emergence of gCW currents in type IIA supergravity (without branes). Following \cite{Heidenreich:2020pkc}, we first recall that given the type IIA field strengths $F_i = d C_{i-1}$, $i = 0,2,4$ and $H_3 = dB_2$, gauge invariant expressions can be constructed
\begin{eqnarray}
    \label{2Aa}
    G_p = F_p - H_3 \wedge C_{p-3} + G_0 \, e^{B_2} \, ,
\end{eqnarray}
which satisfy the following  Bianchi identities and equations of motion 
\begin{eqnarray}
    \label{2Ab}
d G_p - H_3 \wedge G_{p-3}=0\,, ~~~~~
d H_3 = 0\,, ~~~~~~
d \star G_p + H_3 \wedge \star G_{p+2} =0\,. 
\end{eqnarray}
The $G_p$ forms can then be used to obtain an infinite set of higher-form closed currents \cite{Heidenreich:2020pkc}  
\begin{eqnarray}
    \label{2AbA}
 {\mathcal J} = \left\{G^k_0\,,~~ G^k_0 H_3\,, ~~G^k_0 J_4\,, 
~~ G^k_0 J_6\,, ~~G^k_0 J_4^2\,, ~~G^k_0 J_4\wedge J_6\right\} ~~{\rm with}~~ k\geq 0 \, ,
\end{eqnarray}
where $J_4 = G_0 G_4 -\frac12 G_2^2$ and 
$J_6 = G^2_0 G_6 + G_0 G_2 G_4 + \frac13 G^3_2$. 

However, this does not exhaust the whole spectrum of Chern-Weil currents in type IIA supergravity. In fact, in terms of the gravitinos 
$\psi_{L/R}$ and the vielbeins $V^a$, we can define the following cocycles \cite{Cremoninipif}
\begin{equation}
    \label{2Ac}
   \omega^{(2)} = \bar\psi_L \psi_R\,, 
~~~~~~~ \omega^{(3)} =- i (\bar \psi_L \gamma_a \psi_L - \bar\psi_R \gamma_a \psi_R) V^a\,, ~~~~~~
\omega^{(4)} =
(\bar \psi_L \gamma_{ab} \psi_R + 
\bar\psi_R \gamma_{ab} \psi_L) V^a V^b \, ,
\end{equation}
where the first two are on-shell closed, $d \omega^{(2)} =  d \omega^{(3)} =0$, whereas the last one satisfies $d \omega^{(4)} = \omega^{(2)}\wedge \omega^{(3)}$.
Therefore, using $\omega^{(2)}$ and $\omega^{(3)}$ we can build a brand new set of gCW closed currents given by
\begin{eqnarray}
    \label{2Ae}
   \underbrace{\omega^{(2)} \wedge \dots \wedge \omega^{(2)}}_{p-times} \wedge{\mathcal J}\,, \qquad
    \omega^{(3)}  \wedge {\mathcal J}\,, 
\end{eqnarray}
where ${\mathcal J}$ stands for all possible currents in the set \eqref{2AbA} with form degree less or equal to 10. Higher powers of $\omega^{(3)}$ vanish and the product 
$\omega^{(2)}\wedge\omega^{(3)}$ is exact. Therefore, expressions \eqref{2Ae} should cover the complete spectrum of gCW currents in type IIA supergravity. 

A complete analysis of these new currents in the realm of supergravity is beyond the scope of the present contribution. We plan to devote them a deeper study in forthcoming publications.

\section{Super-SymTFT from Supergravity}
\label{sect:general}

A particularly efficient way of describing (generalized) symmetries of a Quantum Field Theory is through the symmetry
topological field theory (SymTFT) \cite{Felder:1999mq,Fuchs:2002cm,Apruzzi:2021nmk} (we refer to \cite{Schafer-Nameki:2023jdn} for an introduction to SymTFT and an exhaustive list of references). This is a BF topological theory in one dimension
higher which encodes generalized symmetries, their defects and their ’t Hooft anomalies. In \cite{Ambrosino:2026hhv}, the concept of SymTFT has been generalized to supersymmetric theories, by embedding the construction in supermanifold. 

Here, we explore the derivation of super-SymTFT topological field theories directly from supergravity or superstring, generalizing similar considerations done in the ordinary non-supersymmetric context (for a short review, see for instance \cite{Schafer-Nameki:2023jdn}). In fact, from \cite{Witten:1998wy,Aharony:1998qu} we learn that in string theory on $AdS_5  \times S^5$, after integrating over the sphere
with $N$ units of 5-form flux, we are left with a topological coupling of the form 
\begin{eqnarray}
\label{doA}
S_{\rm SymTFT} = N \int_{AdS_5} b_2 \wedge d c_2 \, ,
\end{eqnarray}
where $b_2$ and $c_2$ correspond to the NSNS and RR 2-form potentials in Type IIB.
Here, we provide a preliminary discussion on how to reproduce such terms in a  superspace set-up. We focus on a four-dimensional theory with CW and gCW symmetries, whose symTFT lives in five dimensions.

 \subsection{Type IIB}
We first consider the type IIB case. To study the superspace derivation, it is convenient to consider the rheonomic formulation of 
Castellani and Pesando \cite{Castellani:1993ye},  
where the type IIB action is given by 
\begin{eqnarray}
\label{doB}
S_{\rm IIB} = \int_{\mathcal{SM}^{(10|32)}}  \mathcal{L}^{(10|0)} \wedge \mathbb{Y}^{(0|32)} \, .
\end{eqnarray}
Here $ \mathcal{L}^{(10|0)} $ is the rheonomic action \cite{Castellani:1993ye} based on the $(10|32)$ superspace formulation, and $\mathbb{Y}^{(0|32)}$ is the PCO needed to convert the $(10|0)$ form  $ \mathcal{L}^{(10|0)} $ into an integral form. This formulation requires $d  \mathcal{L}^{(10|0)} =0$, in order to make \eqref{doB} independent of the PCO representative. 

If $\mathbb{Y}^{(0|32)}$ respects some isometries, the action will exhibit those isometries manifestly. For example, if the PCO preserves a fraction of supersymmetries, the final action will  reduce to an ordinary superspace integral of a superfield action. 
We choose a PCO of the form
\begin{eqnarray}
\label{doC}
 \mathbb{Y}^{(0|32)} =  \mathbb{Y}^{(0|28)} \wedge \mathbb{Y}^{(0|4)}\,, ~~~~~~
 \mathbb{Y}^{(0|28)} = \prod_{i=1}^{28} \theta_i \delta(d\theta_i)\,, ~~~~~~~
 \end{eqnarray}
where $\mathbb{Y}^{(0|4)}$ is the supersymmetric PCO for the reduced supermanifold $\mathcal{SM}^{(5|4)}$, whereas 
$\mathbb{Y}^{(0|28)}$ projects on the 28 fermions of the supermanifold in a {\it brutal} way, setting them to 
zero, $\theta_i =0$. 

To show how this works concretely, we consider the following suitable set of terms from $\mathcal{L}^{(10|0)}$ 
(see \cite{Castellani:1993ye} in tab. 6 on page 1135)
\begin{eqnarray}
\label{cpA}
&&\hspace{-.8cm}\int_{\mathcal{SM}^{(5|4)}} \int_{\mathcal{SM}^{(5|28)}} \left[ \left(- 560 \, \bar\psi \Gamma^{a_1\dots a_3}\psi V_{a_1} \dots V_{a_3} 
- \frac{14}{3} F^{a_1 \dots a_5} V^{a_6} \dots V^{a_{10}} \epsilon_{a_1 \dots a_{10}} \right. \right. \\
&&\hspace{-.4cm}  \left. \left. + 
\frac{7}{60} \bar \lambda \Gamma^{a_1 \dots a_5} \lambda V_{a_1} \dots V_{a_5}\right)\times
\left( dC_4+ \frac{3i }{8} \epsilon_{IJ} A^I \wedge F^J(A)\right) - 210 i \epsilon_{IJ} 
d A^I  \wedge d A^J \wedge C_4
\right] \wedge \mathbb{Y}^{(0|32)} \nonumber
\end{eqnarray}
where $A^I$ are the $SU(2)/U(1)$ doublet of NSNS-RR 2-form fields and $F^I(A)$ their field strengths. 
$C_4$ is the potential of the RR 4-form and  $F \equiv dC_4+ \frac{3i }{8} \epsilon_{IJ} A^I \wedge F^J(A)$ is the gauge invariant 5-form field strength.
$\psi$ are the gravitinos while $\lambda$ are   
the spin-1/2 fields. 
Now, we integrate the 5-form field strength such that 
\begin{eqnarray}
\label{cpB}
\int_{\mathcal{SM}^{(5|28)}}  F \wedge \mathbb{Y}^{(0|28)} = \frac{N}{2\pi} \, .
\end{eqnarray}
 If we use $\mathbb{Y}^{(0|4)} = \star \omega_{10}^{(5)} = V^a V^b \iota_\psi \gamma_{ab}\iota_\psi \delta^4(\psi)$, where 
 $ \omega_{10}^{(5)} = \epsilon^{abcde} V_a  V_b  V_c \psi \gamma_ {de}\psi $ is the invariant five-dimensional cocycle\footnote{We note that this cocycle is related to 4d cocycles discussed in \cite{Grassi:2025tfp}. In fact, by 
splitting the 5th direction as $V^a = (V^{\hat a}, V^5)$ and using the fact that $\epsilon_{5 \hat a \hat b \hat c \hat d}$ with hatted indices being four-dimensional indices, we find the relation $\omega^{(5|0)} = V^5 (\omega^{(4|0)} + \bar \omega^{(4|0)}) + {\rm  exact~terms}$. }, we obtain
 $\omega_{10}^{(5)} \wedge \mathbb{Y}^{(0|4)} = {\rm Ber}(E)$, 
 where $E$ is the supervielbein $E =(V^a,\psi^\alpha)$, and explicitly
 \begin{equation}
    {\rm Ber}(E)=  \epsilon_{abcde} V^a \dots V^e \delta^{(4)}(\psi) \, .
 \end{equation}

Uunder the flux quantization \eqref{cpB} the first term in \eqref{cpA} reduces to the  three terms in the first bracket integrated on $\mathcal{SM}^{(5|4)}$. However, since we are in five dimensions, the number of integrated vielbeins cannot exceed five. Taking into account that $\mathbb{Y}^{(0|4)}$ contains two vielbeins, 
the second and the third terms of the first bracket are automatically killed. Also the first 
term vanishes because once combined with the PCO it gives rise to ${\rm Tr}(\Gamma^{a_1\dots a_3} \Gamma^{b_1b_2}) =0$. The only surviving term in \eqref{cpA} is then the last one. Integrating over ${\mathcal{SM}}^{(5|28)}$ and exploiting the flux quantization we are left with 
\begin{eqnarray}
\label{cpD}
 210 \, i \, N  \int_{\mathcal{SM}^{(5|4)}}  
\epsilon_{IJ} A^I  \wedge d A^J \wedge \mathbb{Y}^{(0|4)} \, ,
\end{eqnarray}
where $A^I$ are $(2|0)$-superforms, whose general expansion reads
\begin{eqnarray}
\label{cpE}
A^I = A^I_{ab} V^a V^b +  A^I_{a\alpha} V^a  \psi^\alpha + A^I_{\alpha\beta} \psi^\alpha \psi^\beta  \, .
\end{eqnarray}
The  $ A^I_{ab}, A^I_{a\alpha} $ and $A^I_{\alpha\beta} $ components are ordinary superfields. Expressing the $(5|0)$-superform  $\epsilon_{IJ} A^I  \wedge d A^J$ in terms of these components, we expect the multiplication by the $\mathbb{Y}^{(0|4)}$ PCO to select only some components in \eqref{cpD}. Moreover, since $\mathbb{Y}^{(0|4)}$ is manifestly supersymmetric invariant, the final expression \eqref{cpD} will be manifestly supersymmetric, thus expressible as an ordinary Berezin integral of superfields. 

Expression \eqref{cpD} is the five-dimensional version of the general super-BF theory recently used in \cite{Ambrosino:2026hhv} to investigate the super-SymTFT. 
It would be very interesting to write down the integral supermanifold version of this topological field theory. It plays the same role as $N=1$ super Chern-Simons theory in $d=3$ dimension. We should also consider the case $(5|8)$ 
which might play the same role as $N=2$ super-Chern-Simons in $d=3$ \cite{Grassi:2016apf}.

\subsection{Type IIA} 

A similar investigation can be performed in type IIA supergravity. To this purpose one can use the 
rheonomic formulation of $D=11$ supergravity \cite{DAuria:1982uck} with suitable reduction to $(10|32)$ supermanifold. In this process the following term is produced
\begin{eqnarray}
\label{typeIIAaction}
\int_{\mathcal{M}^{(10|32)}} \star F'\wedge F' \wedge \mathbb{Y}^{(0|32)} \, ,
\qquad  \quad F' = F - d B^{(2|0)} \wedge C^{(1|0)} \, , 
\end{eqnarray}
 where $F$ is the field strength of the RR 3-form,  
$B^{(2|0)}$ is the NSNS 2-form and $C^{(1|0)}$ the RR gauge field. Again, we insert the PCO  $\mathbb{Y}^{(0|32)}$ to integrate the $(10|0)$ form on the supermanifold. 

In the present case, we use a different decomposition of the PCO, precisely
\begin{equation}
 \mathbb{Y}^{(0|32)} =  \mathbb{Y}^{(-1|28)} \wedge \omega^{(1|4)}   
\end{equation}
where 
\begin{equation}
\label{cqC}
 \mathbb{Y}^{(-1|28)} = \prod_{i=1}^{28} \theta_i V^a \, \iota_{\psi} \Gamma_a \iota_\psi \,\delta^{(28)}(\psi) \, , \qquad \quad 
\omega^{(1|4)} = V^a V^b V^c \epsilon_{abcd} \, \iota_\psi \Gamma^d \iota_\psi \, \delta^{(4)}(\psi) 
\end{equation}
Now, splitting the integration in \eqref{typeIIAaction} on $\mathcal{M}^{(5|4)} \times \mathcal{M}^{(5|28)}$ and performing the integral on $\mathcal{M}^{(5|28)}$ using  the flux quantization
\begin{eqnarray}
\label{cqD}
\int_{\mathcal{M}^{(5|28)}} \star F' \wedge  \mathbb{Y}^{(-1|28)} =  \frac{N}{2\pi} \, ,
\end{eqnarray}
we land on 
\begin{eqnarray}
\label{cqA}
\int_{\mathcal{M}^{(5|4)}} \omega^{(1|4)} \wedge d C^{(1|0)} \wedge B^{(2|0)} \, .
\end{eqnarray}
This is the structure of a super-BF theory that should correspond to a gCW symmetry in four dimensions generated by the current  $J^{(3|4)} = \omega^{(1|4)} \wedge F^{(2|0)}$, 
where $C^{(1|0)}$ is the background field coupled to $J^{(3|4)}$ and $B^{(2|0)}$ is the background field coupled to 
$F^{(2|4)} = \star F^{(2|0)}$. 

\vskip 10pt
We have provided just a hint on how the symTFT for CW and gCW currents can be obtained directly from supergravity.  The present derivation should  to be made more rigorous and systematic. In particular, we should  explore if there are other terms that might contribute. We leave this interesting project for the future.

\section{Conclusions and Perspectives}
\label{conclusions}

In supersymmetric theories, we have constructed new higher-form global
symmetries associated with higher-superform and higher-integral form
currents. In particular, we have presented a brand new set of conservation laws associated to geometric-super-CW currents (gCW) in super-Maxwell theory and type IIA supergravity. Moreover, in section 7 we have provided a preliminary analysis of a novel derivation of the Super-SymTFT from supergravity. This material is new  and not yet  published.

\noindent
Future projects include:
\begin{itemize}
  \item 
  A better comprehension of the physical implications of the gCW symmetries.

    \item 
     The generalization of our construction to non-abelian theories.
    \item 
    The investigation of non-invertible symmetries in supermanifold.
    \item 
    The generalization of the Super-SymTFT in supermanifold to encode all invertible and non-invertible symmetries and topological phases of a QFT.
    \item 
    A more systematic construction of Super-SymTFT for CW and gCW symmetries from ten dimensional supergravity and string theory. 
\end{itemize}

\vskip 10pt
\noindent
{\bf Acknowledgements}

\noindent
Silvia Penati would like to thank the Organizers of the Cost Action CaLISTA General Meeting 2025, ``Cartan, Generalized and Noncommutative Geometries, Lie Theory and Integrable Systems Meet Vision and Physical Models'' for the kind invitation, and the Corfu Summer Institute for the warm hospitality. P.A. Grassi would like to thank CERN for hospitality during the initial stages of the present project. 
This work was supported in part
by the INFN grants  Gauge and String Theory (GAST) and Gauge Theory, Supergravity, Strings (GSS 2.0). 

\bibliographystyle{unsrt}
\bibliography{super.bib}
\end{document}